\documentclass[prd,amssymb,twocolumn,10pt]{revtex4}
\usepackage{amsmath,amsfonts,amssymb}
\usepackage{color,verbatim,graphicx,float}  

\begin{document}

\title{The Observational Implications of Loop Quantum Cosmology}

\author{Jakub Mielczarek}
\email{jakub.mielczarek@uj.edu.pl}
\affiliation{Astronomical Observatory, Jagiellonian University, 30-244
Cracow, Orla 171, Poland \\
and Laboratoire de Physique Subatomique et de Cosmologie,\\
53, avenue des Martyrs, 38026 Grenoble cedex, France}

\begin{abstract}
In this paper we consider realistic model of inflation embedded in the framework
of loop quantum cosmology. Phase of inflation is preceded here by the phase of a 
quantum bounce. We show how parameters of inflation depend on the initial 
conditions established in the contracting, pre-bounce phase. Our investigations 
indicate that phase of the bounce easily sets proper initial conditions for 
the inflation. Subsequently we study observational effects that might arise due 
to the quantum gravitational modifications. We perform preliminary observational 
constraints for the Barbero-Immirzi parameter $\gamma$, critical density $\rho_{\text{c}}$
and parameter $\lambda$. In the next step we study effects on power
spectrum of perturbations. We calculate spectrum of perturbations from the bounce
and from the joined bounce+inflation phase. Based on these studies we indicate possible way 
to relate quantum cosmological models with the astronomical observations. Using 
the Sachs-Wolfe approximation we calculate spectrum of the super-horizontal 
CMB anisotropies. We show that quantum cosmological effects can, in the natural way,
explain suppression of the low CMB multipoles. We show that fine-tuning is not required
here and model is consistent with observations. We also analyse other possible probes 
of the quantum cosmologies and discuss perspectives of their implementation. 
\end{abstract}

\maketitle

\section{Introduction} \label{sec:intro}
A main obstacle in formulating quantum theory of gravitational interactions
is the lack of any empirical clue. Here the problem is that quantum gravity 
effects are expected to be significant at the energies approaching $10^{19}$ GeV 
(the Planck scale). With the present generation of accelerators, energies up to 
$10^3$ GeV can be reached what is sixteen orders of magnitude below the 
Planck scale. Therefore direct experimental investigation of quantum gravity
effects becomes inaccessible. In other words it is like probing atomic structure 
with Earth size resolution devices. This suggest that alternative methods of 
investigation ought to be taken into account. One, perhaps most promising, possibility
of escape from this impasse are \emph{indirect methods}. In this paper we consider 
one particular type of indirect probing of quantum gravitational effects, which based 
on cosmological observations.

In order to perform any quantitative predictions, mathematical model of the given 
process has to be known. The process considered here is behaviour of the universe 
in the Planck Epoch. In this epoch, evolution of the universe is determined by the 
quantum gravitational effects. In our considerations this phase is described by 
loop quantum cosmology (LQC) \cite{Bojowald:2008zzb}. LQC base on nonperturbative 
approach to quantise gravity called loop quantum gravity (LQG) \cite{Ashtekar:2004eh}.
Starting point in formulating both LQG and LQC is parametrisation of the phase space 
of the gravitational field by $SU(2)$ connection and by its conjugated momenta. These 
canonical fields are so called Ashtekar variables $(A=A^i_a\tau_i dx^a,E=E^a_i\tau^i 
\partial_a)$ which take value in $\mathfrak{su}(2)$ and $\mathfrak{su}(2)^*$ algebras
respectively and fulfil the Poisson bracket
\begin{equation}
\{A^i_a({\bf x}),E^b_j({\bf y})\} = \gamma \kappa \delta^b_a \delta^i_j \delta^{(3)}({\bf x-y}) 
\end{equation}
where $\kappa = 8\pi G$ and  $\gamma$ is Barbero-Immirzi parameter. Parameter $\gamma$ 
is unknown constant of the theory. However, because $\gamma$ is related with a black hole 
entropy, its value can be recovered from comparison with Hawking-Bekenstein formula 
$S_{BH}=\frac{k}{4l^2_{\text{Pl}}}A$. In our considerations we use value $\gamma=\gamma_{\text{M}}
 =0.2375$ calculated by Meissner in Ref. \cite{Meissner:2004ju}.  

Loop quantum cosmology can be considered on the two levels: the first is purely quantum 
approach and the second is semi-classical, effective framework. The first approach is more
complete with respect to effective approach. However the semi-classical approach is more
useful in relating quantum cosmological effects with classical physics. Moreover, main 
features of the complete approach are sufficiently well reproduced by the effective approach.
Because our aim here is to relate effects of LQC with subsequent classical evolution, the 
semi-classical approach will be more adequate. Therefore in all the considerations performed
in this paper we base on semi-classical LQC. 

In the cosmological applications, canonical variables $(A,E)$ can be split for 
the homogeneous and perturbation parts: 
\begin{equation}
A^i_a = \bar{A}^i_a +\delta A^i_a  \ \ \text{and} \ \ E^b_j = \bar{E}^b_j+\delta E^b_j. \label{AEpert}
\end{equation}
In this paper, cosmological background is described by the flat FRW metric, then 
$\bar{A}^i_a =\bar{p} \delta^i_a $ and $\bar{E}^b_j = \gamma \bar{k} \delta ^b_j$.
Here $\bar{k}$ and $\bar{p}$ are new canonical variables which fulfil the Poisson 
bracket $\left\{ \bar{k},\bar{p}\right\}=\frac{\kappa}{3V_0}$. The parameter $V_0$ is 
the fiducial cell which regularise integration over the infinite spatial part. The 
$\bar{p}$ variable can be expressed in terms of the scale factor, $\bar{p}=a^2$ and
$\bar{k}=\dot{\bar{p}}/2\bar{p}$. Having canonical variables one can introduce 
Hamiltonian. The Hamiltonian can be also decomposed for the homogeneous and the 
background parts, $H_\text{G}^{\text{phen}}=\bar{H}_{\text{G}}^{\text{phen}}+\delta 
H_{\text{G}}^{\text{phen}}$. Here the upper index symbolise that classical Hamiltonian 
contains additional phenomenological quantum correction. The lower index indicate that 
the gravitational part is considered. However in the realistic models, the matter 
Hamiltonian also contribute, then $H^{\text{phen}}=H_{\text{G}}^{\text{phen}}+H_{\text{m}}$.
In the considered models there is no quantum corrections to the matter Hamiltonian,
what is indicated by the lack of the upper index. The matter Hamiltonian can be also 
decomposed for the homogeneous and perturbations parts. We will analyse such a case in
this paper. 

In this paper we consider effective LQC with a scalar field. In particular we concentrate
our attention on the model with a massive potential. In this case it will be possible to 
investigate realisation of the inflationary phase. In this approach, parameters of inflation
are dependent on the previous quantum cosmological evolution. We will consider perturbations
in the emerging bounce+inflation scenario. This will allow us to relate quantum cosmological
effects with classical perturbations. This finally let us to study LQC modifications of the CMB 
anisotropy. 

The organisation of the text is the following. In Sec. \ref{Backgroung} we introduce concept 
of cosmic bounce in the framework of the effective LQC. Subsequently in Sec. \ref{QBI}
we construct a model of inflation in the applied framework. We set initial conditions in the 
contracting phase and study how parameters of inflation vary with them. We show that phase 
of bounce can easily set proper initial conditions for the subsequent phase of inflation. 
In the next step, in Sec. \ref{Pert}, we discuss perturbations in the considered model. We 
restrict ourselves to the fluctuations of the scalar field. We calculate spectrum of the 
perturbations from the bounce and from the joined bounce+inflation phase. These results can 
be applied in calculating spectrum of the CMB anisotropies. In Sec. \ref{CMB}, based on 
Sachs-Wolfe approximation, we calculate spectrum of temperature anisotropies in CMB. We 
show that phase of bounce can lead to suppression of the low multipoles in the spectrum 
of CMB anisotropies.  Subsequently, in Sec. \ref{QGProbes}, we discuss other kinds of the 
observational probes of LQC.  Finally in Sec. \ref{Summary} we summarise our results.
 
\section{Background Dynamics} \label{Backgroung}

In our considerations, Hamiltonian of the gravitational homogeneous part 
is given by  
\begin{equation}
\bar{H}_{G}^{\text{phen}} = -\frac{3V_0}{\kappa} \sqrt{\bar{p}}
 \left( \frac{\sin \bar{\mu}\gamma \bar{k} }{\bar{\mu}\gamma} \right)^2. 
\end{equation}
The crucial element of this Hamiltonian is the factor $\bar{\mu}$. This parameter 
contains details of the quantum modifications and when $\bar{\mu}\rightarrow 0$,
the classical limit is recovered. The main ambiguity in LQC comes from the choice 
of $\bar{\mu}$.  The mostly used form of $\bar{\mu}$ is this given by $\bar{\mu}=
\sqrt{\frac{\Delta}{\bar{p}}}$ where $\Delta=2\sqrt{3}\pi \gamma l^2_{\text{Pl}}$. 
In our investigations we use this particular expression for $\bar{\mu}$. The choice of $\Delta$ is 
motivated by the existence of the gap in the spectrum of area operator in LQG. 
However, $\Delta$ is the result of the kinematic sector of LQG and extrapolation 
of this result to LQC is an assumption. This issue is discussed in Ref.
\cite{Dzierzak:2008dy,Bojowald:2008ik}. The problem here is that 
the relation between LQC and LQG is not fully understood. In particular it should 
be possible to derive LQC directly from LQG and then problem of ambiguities in LQC 
should be overcame. Recently one promising step has been done towards to derive 
LQC from LQG. In their work, Rovelli and Vidotto show how LQC can be derived in
the spin networks formalism. Another possibility is those presented in 
\cite{Malkiewicz:2009zd,Dzierzak:2008dy} where $\bar{\mu}=\lambda/\sqrt{\bar{p}}$ and $\lambda$ 
is some unknown constant unrelated with $\Delta$. From this point of view $\lambda$ 
is some phenomenological parameter which should be determined from the observations 
rather from the theory.

\begin{figure}[ht!]
\centering
\includegraphics[width=8cm,angle=0]{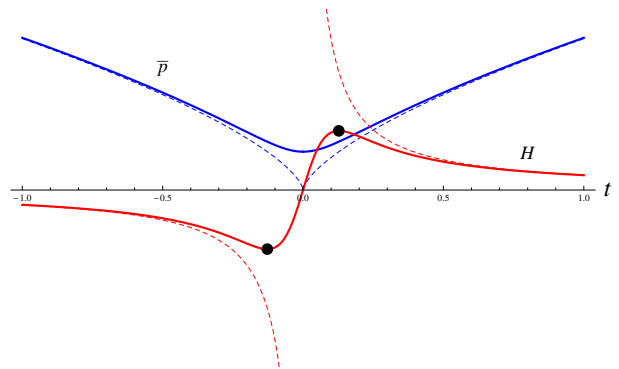}
\caption{Evolution of variable $\bar{p}$ and Hubble factor $H$ in the bouncing 
universe (thick lines) with a free scalar field. Dashed lines represents classical 
evolution. Dots represents  the points $\left(t_{\pm},\mp H_{\text{max}} \right)=\left( \pm \frac{1}{\sqrt{3\kappa\rho_{\text{c}}}},\mp \sqrt{\frac{\kappa}{12} \rho_{\text{c}}}\right)$. }
\label{Bounce}
\end{figure}

Taking Hamilton equation $\dot{\bar{p}}=\{\bar{p},\bar{H}_{m}+\bar{H}_{\text{G}}^{\text{phen}}\}$ 
together with the scalar constraint $\bar{H}_{m}+\bar{H}_{G}^{\text{phen}} = 0$ one can derive
modified Friedmann equation
\begin{equation}
H^2 := \left(\frac{1}{2\bar{p}} \frac{d\bar{p}}{dt}\right)^2 = \frac{\kappa}{3}\rho 
\left(1-\frac{\rho}{\rho_c}\right)  
\label{Heff}
\end{equation}
where 
\begin{equation}
\rho_c = \frac{\sqrt{3}}{16\pi^2 \gamma^3 l^4_{\text{Pl}}} =  0.82\ m^4_{\text{Pl}} \label{rhoc}
\end{equation}
is critical energy density. As one can find from Eq. (\ref{Heff}) the physical solutions 
are allowed only for $\rho\leq \rho_c$. The $\rho=\rho_c$ is a turning point, commonly 
called a bounce. Moreover, while $\rho \rightarrow 0 $ the classical dynamics is recovered.
One can also find that maximal value of the Hubble factor defined in Eq. (\ref{Heff}) is 
reached for $\rho_{\text{max}}=\frac{\rho_{\text{c}}}{2}$. The maximal value of the Hubble 
factor is $H^2_{\text{max}}=\frac{\kappa}{12} \rho_{\text{c}}$. In Fig. \ref{Bounce} we 
show typical symmetric bounce obtained for the model with a free scalar field.

\section{Quantum Bounce and Inflation} \label{QBI}

In this section we are going to construct realistic model of inflation embedded
in the framework of effective loop quantum cosmology. Qualitative studies of inflation
in LQC have been already performed in Ref. \cite{Singh:2006im}. Issue of inflation in LQC 
has been raised also in Ref.  \cite{Artymowski:2008sc,Zhang:2007bi}. In our studies we 
model phase of inflation with the massive scalar field. Since we consider flat FRW model 
the equation for the homogeneous component of the field $\varphi$ holds the classical form
\begin{equation}
\frac{d^2\bar{\varphi}}{dt^2}+3H\frac{d\bar{\varphi}}{dt}+\frac{dV}{d\bar{\varphi}} = 0 \label{backfield}
\end{equation}
where massive potential is given by
\begin{equation}
V(\varphi)= \frac{m^2}{2}\varphi^2.
\end{equation}
Energy density of the considered homogeneous scalar field is 
\begin{equation}
\rho = \frac{1}{2}\left( \frac{d\bar{\varphi}}{dt}\right)^2 +V(\bar{\varphi}).
\end{equation}
 
Dynamics of the model is governed by the set of equations
\begin{eqnarray}
\frac{d H}{d t} &=& \kappa \frac{P^2}{2}
\left[\frac{2}{\rho_{\text{c}}} \left(\frac{P^2}{2}+V(\bar{\varphi})\right)-1  \right],
 \label{ds1}  \\
\frac{d\bar{p}}{dt} &=& 2H\bar{p},  \label{ds2}  \\
\frac{d\bar{\varphi}}{dt} &=& P,      \label{ds3}      \\
\frac{dP}{dt} &=& -3HP  -\frac{dV(\bar{\varphi})}{d\bar{\varphi}}.   \label{ds4} 
\end{eqnarray}

Phase space of this system has been studied in Ref. \cite{Singh:2006im}. Analogous dynamics 
for the closed FRW model has been studied recently in Ref. \cite{Mielczarek:2009kh}.

In our consideration we are going to restrict ourselves to the subset of initial conditions.
We consider initial condition in the pre-bounce stage at the time $t_0$. We make very general 
assumption that  at this arbitrary time, field is placed in the bottom of the potential, 
therefore $\bar{\varphi}(t_0)=0$. Since $\rho\leq \rho_c$ we obtain restriction for $P$ at $t_0$, namely  
$|P(t_0)| \leq \sqrt{2 \rho_{c}}$. Taking particular value of $P(t_0)$ we can compute value of 
the Hubble factor  
\begin{equation}
H(t_0)=-\sqrt{\frac{\kappa}{3} \frac{P^2(t_0)}{2}\left(1-\frac{P^2(t_0)}{2\rho_c} \right) }. \label{Ht0}
\end{equation}

In the dynamical system defined by Eq. \ref{ds1}-\ref{ds4}, one can distinguish
subsystem ($H$, $\bar{\varphi}$, $P$) whose evolution does not depend on $\bar{p}$. 
In this subsystem, initial conditions are specified by the value of $P(t_0)$, 
because $\bar{\varphi}(t_0)=0$ and $H(t_0)$ is given by Eq. \ref{Ht0}. Initial 
value of $\bar{p}$ can be set arbitrary since only changes of $\bar{p}$ have 
physical meaning. 

In the subsequent part of this section we will show how parameters of inflation 
depend on the choice of $P(t_0)$ and $m$.

\subsection{Analytical approximations}

\begin{figure}[ht!]
\centering  
\includegraphics[width=7cm,angle=0]{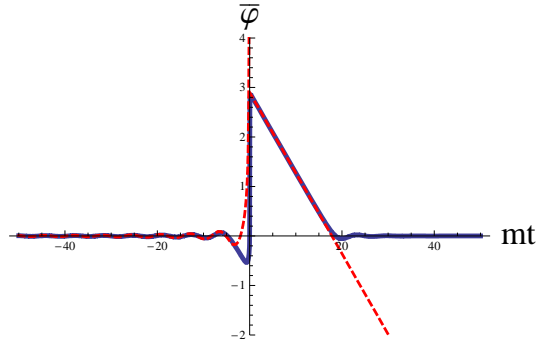} 
\caption{Evolution of the field $\bar{\varphi}$. Dashed lines
represents analytical approximations. Here $m=10^{-4}m_{\text{Pl}}$ and 
$\bar{\varphi}_{\text{max}}=2.9m_{\text{Pl}}$.}
\label{varphi}
\end{figure}

\begin{figure}[ht!]
\centering  
\includegraphics[width=7cm,angle=0]{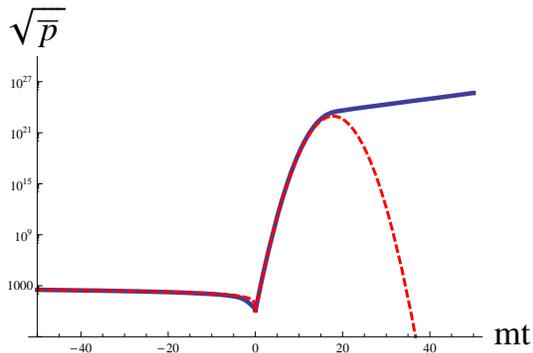} 
\caption{Evolution of the variable $\bar{p}$. Dashed lines
represents analytical approximations. Here $m=10^{-4}m_{\text{Pl}}$ and 
$\bar{\varphi}_{\text{max}}=2.9m_{\text{Pl}}$.}
\label{sqrtpbar}
\end{figure}

Dynamics of the considered model is nonlinear and cannot be traced analytically.
However we can distinguish two regions where approximated analytical solutions 
can be found. The first is the phase of contraction and the second is the phase 
of slow-roll inflation. In this first phase, field oscillate in the bottom of 
the potential well. Therefore value of $\bar{\varphi}=0$ is reached many times, 
what motivate our choice of the initial condition  $\bar{\varphi}(t_0)=0$.
The oscillations are amplified when the moment of the bounce 
is approached. In this regime evolution of the field is approximated by 
\begin{equation}
\bar{\varphi}(t)= C \frac{\cos{[m(t-t_0)]}}{\bar{p}^{3/4}}.  \label{oscilphi}
\end{equation}
In the subsequent phase of the slow-roll inflation, evolution of the scalar field is 
approximated by
\begin{equation}
\bar{\varphi}(t)=\bar{\varphi}_{\text{max}}-\frac{m}{\sqrt{12\pi G}} t.
\end{equation}
In Fig. \ref{varphi} we show exemplary evolution of the 
scalar field for the considered model. We show also how approximated solutions fit to the
solution obtained numerically.  
In the contracting phase, scalar field follows the approximated 
solution given by Eq. \ref{oscilphi}. Therefore energy density 
behaves like $\rho\simeq \frac{C^2m^2}{2\bar{p}^{3/2}}$. This is equivalent 
with the case of the universe filled by the dust matter. Corresponding
evolution of the parameter $\bar{p}$ is in this case given by
\begin{equation}
\bar{p}(t) =\left( -\frac{\sqrt{3\kappa}}{2}Cm t+\bar{p}_i^{3/4} \right)^{4/3}. 
\end{equation}
In the subsequent phase of the slow-roll inflation, the approximated solution
is given by
\begin{equation}
\bar{p}(t) = \bar{p}_i \exp \left[ \sqrt{\frac{16\pi G}{3}}m\left(\bar{\varphi}_{\text{max}} 
t -\frac{m}{\sqrt{48\pi G}} t^2  \right)\right].  
\end{equation}
In Fig. \ref{sqrtpbar} we show exemplary evolution of the canonical variable $\bar{p}$ 
for the considered model. We show also how approximated solutions fit to the
solution obtained numerically.  

\subsection{Conditions for inflation}

When the field $\bar{\varphi}$ reaches a point of maximal displacement 
$\bar{\varphi}_{\text{max}}$ then turns back and consequently 
$P(\bar{\varphi}_{\text{max}})=0$. At this point energy of the field is 
given only by the potential part. Because the total energy density is 
restricted $\rho \leq \rho_{\text{c}}$, we obtain the following constraint  
\begin{equation}
\left| \bar{\varphi}_{\text{max}}\right|  \leq \frac{\sqrt{2\rho_c}}{m} = 
1.3 \frac{m^2_{\text{Pl}}}{m}. \label{phimaxbound}
\end{equation}

The parameter $\bar{\varphi}_{\text{max}}$  is important since 
gives good characterisation of the inflation. Moreover based on its
value one can express e-folding number as follows
\begin{equation}
N \simeq  2\pi \frac{\bar{\varphi}^2_{\text{max}} }{m^2_{\text{Pl}}}. \label{Ndef}
\end{equation}
Based on this expression and Eq. \ref{phimaxbound} we obtain another bound 
\begin{equation}
N m^2  \leq  \frac{4\pi \rho_c}{m^2_{\text{Pl}}} = 10.3 \ m^2_{\text{Pl}}. \label{Nmbound}
\end{equation}

In the bounds given by Eq. \ref{phimaxbound} and Eq. \ref{Nmbound} we have assumed
value of $\rho_{\text{c}}$ given by Eq. \ref{rhoc}. However these bounds can be seen 
also in the different way. Namely, having parameters of inflation one can restrict the 
value of $\rho_{\text{c}}$.  

In the considered setup the value of $\bar{\varphi}_{\text{max}}$  depends only on $P(t_0)$ and $m$.
It is worth to study how the value of $\bar{\varphi}_{\text{max}}$ is sensitive on them. The results of 
our investigation are shown in Tab. \ref{Table}.
\begin{table}[!h]
\begin{tabular}{|c|ccccccc|}
\hline
$P(t_0)\diagup m$  & 1 &  $10^{-1}$ &  $10^{-2}$ &  $10^{-3}$ &  $10^{-4}$ &  $10^{-5}$ &  $10^{-6}$  \\ 
\hline
1 & 0.5 & 0.8 & 1.1 & 1.4 & 1.8 & 2.1 & 2.2 \\
$10^{-1}$ & {\color{red} 0.9 }& 1.1 & 1.5 & 1.8 & 2.2 & 2.4 & 2.7 \\
$10^{-2}$ & 0.7 & {\color{red} 1.6 }& 1.8 & 2.2 & 2.5 & 2.8 & 3.0 \\
$10^{-3}$ & & 1.3 & {\color{red}2.2} & 2.5 & 2.9 & 3.2 &{\color{blue} 3.4 }\\
$10^{-4}$ & &  & 2.0 & {\color{red}3.0 }& 3.2 & 3.6 & 4.0 \\
$10^{-5}$ & &  &  & 2.7 & {\color{red}3.7} & 3.9 & 4.2 \\
$10^{-6}$ & &  &  &  & 3.4 & {\color{red}4.4} & 4.5 \\
$10^{-7}$ & &  &  &  &  & 4.1 & {\color{red}5.0} \\
\hline
\end{tabular} 
\caption{Values of $\bar{\varphi}_{\text{max}}$ for the different $P(t_0)$ and $m$ 
(all parameters in Planck units).}
\label{Table}
\end{table}

In this table we collected values of  $\bar{\varphi}_{\text{max}}$ obtained for the different values
of initial parameters. The main conclusion coming from this data is that despite the substantial
change of the parameters, the value of $\bar{\varphi}_{\text{max}}$  does not change considerably. 
Therefor no fine-tuning is required to obtain proper phase of inflation. Moreover 
it is worth to stress that phase of bounce indeed leads to the proper inflationary scenario.
In the classical considerations the high value of $\bar{\varphi}_{\text{max}}$  has to be assumed, 
while in the LQC inspired model, this value can be obtained naturally. 

\subsection{Constraining $\rho_{\text{c}}$, $\gamma$ and $\lambda$}

Now let us assume that parameters of inflation are given by $m=10^{-6} m_{\text{Pl}}$ and 
$\bar{\varphi}_{\text{max}}=3.4 m_{\text{Pl}}$, what gives $N \simeq 73$. These values comes 
partially from the CMB observations and partially from the the requirement of solving the 
horizon problem. Based on these values and from Eq. \ref{phimaxbound} we obtain the 
following constraint 
\begin{equation}
\rho_{\text{c}}  \geq 6 \cdot 10^{-12}  m^4_{\text{Pl}}. \label{rhocbound}
\end{equation}
This constraint is not very useful because is very week. However the point 
is just to show possibility of constraining and indicate the presently available 
cosmological bounds. Lets us also examine restriction of the Barbero-Immirzi 
parameter. Taking  Eq. \ref{rhoc} together with the constraint Eq. \ref{rhocbound}
we obtain
\begin{equation}
\gamma \leq 1222.  
\end{equation}

Now one can also constraint the value of parameter $\lambda$ from the formulation 
presented in Ref. \cite{Malkiewicz:2009zd}, where $\bar{\mu}=\lambda/\sqrt{\bar{p}}$.
Here $\lambda$ is phenomenological scale of the loop quantisation (polymerisation).
Now $\rho_\text{c} = \frac{3}{8\pi} \frac{m^2_{\text{Pl}}}{\gamma^2\lambda^2}$ and 
assuming $\gamma=\gamma_{\text{M}}=0.2375$ we obtain 
\begin{equation}
\lambda \leq 7 \cdot 10^4 l_{\text{Pl}}.  
\end{equation}

\section{Perturbations} \label{Pert}

Cosmological perturbations are essential element in searching for the 
quantum gravitational signatures. It is because the super-horizontal 
modes of perturbations can carry information from the inflationary 
and even pre-inflationary epoch. Another important issue is that 
generation of perturbations can have quantum gravitational origin.

In LQC, as we already mentioned in Introduction, the perturbations are 
introduced according to Eq. \ref{AEpert}. Perturbations $(\delta A,\delta E)$ 
can be split for the: 
\begin{itemize}
\item  \textbf{Scalar modes} (coupled with a scalar field) 

This type of perturbations with LQG corrections was studied in Ref. 
\cite{Bojowald:2006tm,Bojowald:2008jv,Bojowald:2008gz}. However until not
only inverse volume corrections have been introduced systematically. Consistent
introduction of the holonomy corrections to the scalar modes is still awaiting.  

\item \textbf{Vector modes}

This kind of perturbations are of the secondary importance in cosmology.
It is because they are decaying modes and cannot affect the CMB substantially.
Nevertheless this type of perturbations was studied in the context of LQC in
Ref. \cite{Bojowald:2007hv}.

\item \textbf{Tensor modes} (gravitational waves)

Tensor modes in LQC were studied in numerous papers. In contrast to the other two types 
of perturbations, in this case also phenomenological implications were studied. The effect of inverse volume 
corrections were studied in Ref. \cite{Bojowald:2007cd, Copeland:2008kz, Grain:2009cj}  
while effects of the holonomy corrections were 
investigated in Ref. \cite{Bojowald:2007cd,Grain:2009kw, Mielczarek:2009vi,Barrau:2008hi, Mielczarek:2008pf}. 
In these papers, creation of gravitational waves was considered either during the phase of a bounce or 
during the phase of inflation. The next natural step here is to consider creation of the 
gravitational waves at the joined bounce+inflation phase considered in Sec. \ref{QBI}.

\end{itemize}

I this section we will consider simplified model of perturbations. 
Namely we will consider perturbations of the matter field only. The
gravitational field is set to be homogeneous. This is only idealisation,
however many results from these studies can be extrapolated to the case 
of scalar and tensor perturbations. 

\subsection{Scalar field perturbations}

Hamiltonian of the scalar field is given by 
\begin{eqnarray}
{H}_{\varphi} &=& \int_{V_0} d^3{\bf x} \, \left( \frac{1}{2}\frac{\pi^2_{\varphi}}{\sqrt{|\det E|}}
\right. \nonumber \\  &+& \left. \frac{1}{2} \frac{E^a_i E^b_i \partial_a \varphi \partial_b \varphi  }{\sqrt{|\det E|}} 
+\sqrt{|\det E|} V(\varphi) \right).
\end{eqnarray}
Similarly like in the case of gravitational field, scalar field can be decomposed for homogeneous and
perturbation parts
\begin{eqnarray}
\varphi = \bar{\varphi}+\delta\varphi \ \ \pi_{\varphi} = \bar{\pi}_{\varphi}+\delta\pi_{\varphi}. 
\end{eqnarray}
Here homogeneous parts are defined as follows
\begin{eqnarray}
\bar{\varphi} (t) &=& \frac{1}{V_0}\int_{V_0} d^3{\bf x}  \varphi ({\bf x},t), \\  
\bar{\pi}_{\varphi}(t) &=& \frac{1}{V_0}\int_{V_0} d^3{\bf x}  \pi_{\varphi} ({\bf x},t).
\end{eqnarray}

Equations of motion for the background and perturbation parts are given by
\begin{eqnarray}
\dot{\bar{\varphi}} &=& \{\bar{\varphi},H_{\varphi}\} =  \bar{p}^{-3/2}\bar{\pi}_{\varphi}, \label{pert1}     \\
\dot{\bar{\pi}}_{\varphi} &=& \{\bar{\pi}_{\varphi},H_{\varphi}\} = 
-\bar{p}^{3/2} \frac{dV(\bar{\varphi})}{d\bar{\varphi}},  \label{pert2}  \\
\delta\dot{\varphi} &=& \{\delta\varphi,H_{\varphi}\} =   \bar{p}^{-3/2}  \delta\pi_{\varphi},  \label{pert3}     \\
\delta\dot{\pi}_{\varphi} &=& \{\delta\pi_{\varphi},H_{\varphi}\} = 
\left( \sqrt{\bar{p}}\nabla^2  \delta\varphi \right. \nonumber   \\    
&-& \left. \bar{p}^{3/2} \frac{d^2V(\bar{\varphi})}{d\bar{\varphi}^2} \delta\varphi \right). \label{pert4} 
\end{eqnarray}
Combining equations (\ref{pert1}) and (\ref{pert2})  we obtain equation (\ref{backfield}).
Variable $\delta\varphi$ can be decomposed for the Fourier modes
\begin{equation}
\delta\varphi(\eta,{\bf x}) = \int \frac{d^3{\bf k}}{(2\pi)^3}  
\frac{u(\eta,{\bf k})}{\sqrt{\bar{p}}} e^{i{\bf k}\cdot {\bf x}}. 
\end{equation}
Based on this decomposition and on the equations (\ref{pert3}) and (\ref{pert4}) we obtain equation
\begin{equation}
\frac{d^2}{d\eta^2} u(\eta,{\bf k}) +[k^2 +m^2_{\text{eff}}] u(\eta,{\bf k}) = 0, \label{modeeq}
\end{equation}
where $k^2={\bf k}\cdot {\bf k}$ and 
\begin{equation}
m^2_{\text{eff}}  = \bar{p} \frac{d^2V(\bar{\varphi})}{d\bar{\varphi}^2}  - \frac{1}{\sqrt{\bar{p}}}\frac{d^2\sqrt{\bar{p}}}{d\eta^2}.  
\end{equation}

In order to describe properties of the perturbations it is useful to introduce 
quantity called power spectrum which is defined as follows
\begin{equation}
\mathcal{P}_u = \frac{k^3}{2\pi^2} |u|^2. \label{DefPowerSpect}
\end{equation}

In the following two subsections we compute this quantity for two 
quantum cosmological models. The first will be the model of the 
symmetric bounce with the free scalar field. The second will be the 
model with the joined bounce+inflation phase. 

\subsection{Symmetric bounce} 

\begin{figure}[ht!]
\centering
\includegraphics[width=8cm,angle=0]{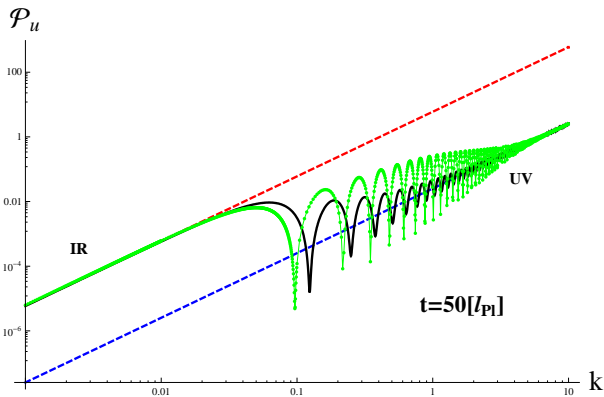}
\caption{Numerical power spectrum of the field $u$ (green points). Black line represent the 
analytical spectrum given by Eq. \ref{PowSpectu} with $U_0=3 m^2_{\text{Pl}}$ and $\eta_0 = 0.1$. 
Dashed lines represents UV and IR limits given by Eq. \ref{UV} and Eq. \ref{IR}.}
\label{PowSpectuFig}
\end{figure}

As the first case we consider scalar field perturbations at the 
symmetric bounce. In the considered case the field is free, $V=0$.
We set the initial conditions to be the Minkowski vacuum
\begin{equation}
u_{\text{in}} = \frac{e^{-ik\eta}}{\sqrt{2k}}. \label{SubHorSol1}
\end{equation}
This approximation works however only for the sub-horizontal modes. 
With these initial conditions we solve the equation (\ref{modeeq})
numerically. Based on these computations we obtain the power spectrum 
of the field $u$ in the post-bounce phase. We show the results in 
Fig. \ref{PowSpectuFig}.
In this figure, black straight line represent analytical approximation of the spectrum.
In order to derive this approximation we assume
\begin{equation}
u_{\text{out}} = \frac{\alpha_k}{\sqrt{2k}}e^{-ik\eta}+\frac{\beta_k}{\sqrt{2k}}e^{ik\eta}.
\end{equation}
Here the relation $|\alpha_k|^2-|\beta_k|^2=1$ holds, as a consequence of the normalisation condition.
Now we base on the integral representation
\begin{equation}
 u(\eta,{\bf k}) =  u_{\text{in}}(\eta,{\bf k})+\frac{1}{k} \int_{-\infty}^{\eta}d\eta' U(\eta') 
\sin k (\eta-\eta') u(\eta,{\bf k})
\end{equation}
of the  Eq. \ref{modeeq} where $U(\eta)=-m^2_{\text{eff}}(\eta)$. Solving this equation in the 
first order of perturbative expansion we compute values of $\alpha_k$ and $\beta_k$. 
Performing approximation 
$U(\eta)=U_0\Theta(\eta-\eta_0)\Theta(\eta_0-\eta)$ where 
\begin{equation}
U_0 :=-m^2_{\text{eff}}(t=0) = \frac{\kappa}{3} \left( 2 \bar{\pi}_{\varphi} \rho_c \right)^{2/3}
\end{equation}
we find
\begin{eqnarray}
\alpha_k\approx1+\frac{i}{2k} \int_{-\infty}^{\infty} d\eta U(\eta)= 1+i\frac{U_0\eta_0}{k},  \\
\beta_k\approx-\frac{i}{2k} \int_{-\infty}^{\infty} d\eta U(\eta)e^{-2ik\eta}=-i\frac{U_0}{2k^2}\sin(2k\eta_0).
\end{eqnarray}
Now with use of definition of the power spectrum we obtain
\begin{widetext}
\begin{equation}
\mathcal{P}_u=\left( \frac{k}{2\pi}\right)^2+\frac{U_0 \left(\sin \left[2 k \eta _0\right]{}^2 U_0+
4 k^2 U_0 \eta _0^2+4 k \sin \left[2 k \eta _0\right] \left(k \sin[2 \eta k]-
\cos[2 \eta k] U_0 \eta _0\right)\right)}{16 \pi k^2} \label{PowSpectu}
\end{equation}
\end{widetext}
In the UV and IR limits, the power spectrum  given by Eq. \ref{PowSpectu} behaves like
\begin{eqnarray}
\mathcal{P}_u^{\text{UV}} &\rightarrow& \left( \frac{k}{2\pi}\right)^2, \label{UV} \\
\mathcal{P}_u^{\text{IR}} &\rightarrow& \left( \frac{k}{2\pi}\right)^2 
\left(1+4U_0\eta_0\eta+4U_0^2\eta_0^2\eta^2 \right). \label{IR} 
\end{eqnarray}
The term in the second bracket in Eq. \ref{IR} is the difference between 
the UV and IR slopes in Fig. \ref{PowSpectu}. 

The spectrum obtained in this subsection is similar to this obtained in the case 
of gravitational waves in Ref. \cite{Mielczarek:2009vi}.  The difference is that 
in the case of the scalar field the effective mass $m^2_{\text{eff}}$ is negative 
in vicinity of the bounce while in case of the gravitational waves with holonomy corrections,
 the effective mass is positive function during the whole evolution.

\subsection{Bounce+Inflation model}

\begin{figure}[ht!]
\centering
\includegraphics[width=8cm,angle=0]{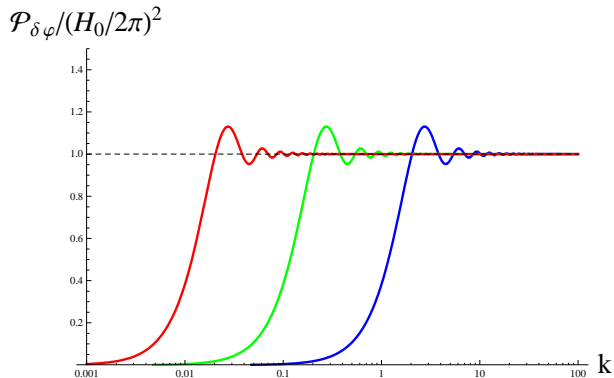}
\caption{Power spectrum of the field $\delta \varphi$. Here $\eta_i=-1, -10, -100$ (from right to left).}
\label{PSBI}
\end{figure}

Phase of symmetric bounce is very idealised situation. More physically relevant is
the joined bounce+inflation phase. In this subsection we show simple analytical model
of perturbations in this phase. It will be a model of perturbations created in the 
scenario described in Sec. \ref{QBI}. In the contracting phase the sub-horizontal modes are 
given by Eq. \ref{SubHorSol1}. The subsequent phase of inflation is approximated by 
de Sitter phase where evolution of the scale factor is given by $a=-\frac{1}{H_0\eta}$.
In this phase modes of fluctuations are given by superposition of Bunch-Davies 
modes
\begin{equation}
u_{\text{out}} = \frac{\alpha_k}{\sqrt{2k}}e^{-ik\eta}\left(1-\frac{i}{k\eta} \right)+
\frac{\beta_k}{\sqrt{2k}}e^{ik\eta}\left(1+\frac{i}{k\eta} \right)
\end{equation}
where relation $|\alpha_k|^2-|\beta_k|^2=1$ holds.
Performing matching conditions $u_{\text{in}}(\eta_i)=u_{\text{out}}(\eta_i)$ and 
$u'_{\text{in}}(\eta_i)=u'_{\text{out}}(\eta_i)$ we determinate coefficients 
\begin{eqnarray}
\alpha_k&=&-\frac{1-2 i k \eta_i -2 k^2 \eta_i^2}{2 k^2 \eta_i^2},    \\
\beta_k&=&-\frac{e^{-2 i k \eta_i }}{2 k^2 \eta_i ^2}. \\
\end{eqnarray}
Based on this we can derive the power spectrum. We show plot of this spectrum in Fig. \ref{PSBI}.

The obtained spectrum is characterised by the suppression for the low values of $k$. 
For the large $k$ the spectrum holds the inflationary form. Another important feature are 
oscillations which are residue of the bouncing phase. We see that damping begins 
when $-\eta_i k \sim 1$. This corresponds to the $k$ on horizon scale at time $\eta_i$. 
Based on this we define $k_* = -\frac{1}{\eta_i}$. At the time $\eta_i$ a value of the 
scale factor is given by $a_i = -\frac{1}{H_0\eta_i}$, therefore $k_* = a_i H_0$. Defining 
the length scale $\lambda_* = a_i/k_*$ we obtain $\lambda_*=\frac{1}{H_0}$. Today value of 
the $\lambda_*$ is given by  $\lambda_* a_0/a_i$ where $a_0$ is the present value of a scale factor.

The similar power spectrum to this obtained here was derived also in Ref. \cite{Contaldi:2003zv,Piao:2003zm}. 

\section{CMB Anisotropy} \label{CMB}

As we have shown, there are two effects of the bounce phase on the primordial 
power spectrum: damping of the low energy modes and oscillations. This first 
effect is dominant and in this section we are going to investigate its impact 
of the CMB anisotropy. 

\subsection{Sachs-Wolfe approximation}

\begin{figure}[ht!]
\centering
\includegraphics[width=8cm,angle=0]{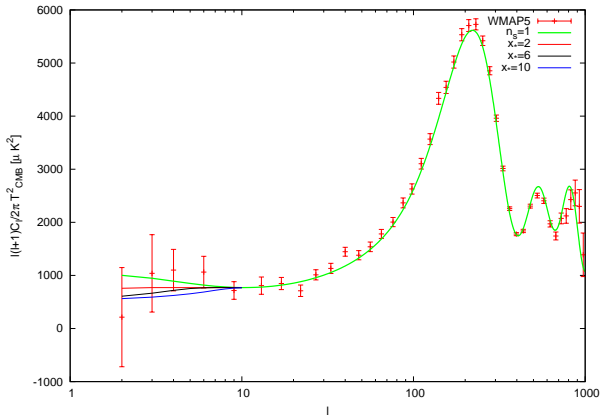}
\caption{Spectrum of CMB anisotropy.}
\label{CMBSpect}
\end{figure}

Because we expect that effects of the bounce can affect super horizontal modes, the 
Sachs-Wolfe approximation can be used to study resulting spectrum of CMB. In 
this approximation sub-horizontal evolution of the primordial plasma is neglected 
since it does not affect the considered modes. Expression for the CMB multipoles is 
given by
\begin{equation}
C_l = \frac{4\pi}{25} \int_0^{\infty} \frac{dk}{k} \mathcal{P}_{\mathcal{R}}(k) j^2_l(kD_{\star}) \label{SW}
\end{equation}
where $D_{\star} = 1.4 \cdot 10^4$ Mpc is distance to last scattering shell. Moreover $\mathcal{P}_{\mathcal{R}}(k)  \equiv  \frac{k^3}{2\pi^2}|\mathcal{R}_{\bf k}|^2$ where $\mathcal{R} = - \frac{v}{z}$ 
 and $v$ is Mukhanov variable which fulfils equation
\begin{equation}
v^{''}+\left[k^2-\frac{z^{''}}{z}\right]v=0  \ \text{and where} \ z = \frac{\sqrt{\bar{p}}\dot{\bar{\varphi}}}{H}.\nonumber
\end{equation}
We see that while approximation $z^{''}/z \approx a^{''}/a$ holds, then evolution
of $v$ and $u$ variables are the same. The validity of this approximation was indicated
in  Ref.  \cite{Contaldi:2003zv,Piao:2003zm}. The assumption we made here is lack of 
the quantum modification to the equation for the $v$ variable.  However we  
do not have a right equation for the scalar modes in presence of holonomy corrections. 
In case of the tensor modes, it was shown in Ref. \cite{Mielczarek:2009vi} that corrections 
to the mode equation do not change the spectrum significantly. It was shown that the 
shape of the spectrum is determined mainly by the background evolution. 
Therefore we assume here that the spectrum of the scalar perturbations is not affected 
significantly by the holonomy corrections to the mode equation. Then spectrum from the
joined bounce and inflation phase should have generic form derived in Ref. \ref{Pert}.
In order to build the analytic model we can average the spectrum over the substandard oscillations.
Then the power spectrum from the joined bounce+inflation is approximated by 
\begin{equation}
\mathcal{P}_{\mathcal{R}}(k) = 
\mathcal{A}^2_{\mathcal{R}}\Theta(k-k_*)+\mathcal{A}^2_{\mathcal{R}}\Theta(k_*-k)\left( \frac{k}{k_*}\right)^2  \label{PR}
\end{equation}
Based on this spectrum one can derive analytical formula for the spectrum of the low multipoles 
of the CMB anisotropy. Instead of the variable $C_l$ it is convenient to consider variable
\begin{equation}
\mathcal{C}_l \equiv \frac{l(l+1)}{2\pi} C_l. 
\end{equation}
It is motivated by the fact that for the scale invariant 
Harrison-Zeldovich power spectrum, this quantity holds constant value. 
Performing integral (\ref{SW}) with the spectrum (\ref{PR}) we obtain
\begin{equation}
\mathcal{C}_l = \mathcal{C}_l^{\text{inflation}}+\mathcal{C}_l^{\text{bounce}}
\end{equation}
where 
\begin{equation}
\mathcal{C}_l^{\text{inflation}} = \frac{\mathcal{A}^2_{\mathcal{R}}}{25}
\end{equation}
and
\begin{widetext}
\begin{equation}
\mathcal{C}_l^{\text{bounce}}=\frac{\mathcal{A}^2_{\mathcal{R}}}{25} \frac{x^2_*}{4^{1+l}\Gamma^2(l+3/2)} \left[
l{_2F_3}(1+l,1+l;l+3/2,2+l,2l+2;-x_*^2)-(1+l){_1F_2}(l;l+3/2,2l+2;-x_*^2)\right].
\end{equation}
\end{widetext}
Here we have introduced parameter $x_*=k_*D_{\star}$. In Fig. \ref{CMBSpect} we show
$\mathcal{C}_l$ spectrum with parameter $\mathcal{A}^2_{\mathcal{R}}=2.6 \cdot 10^{-9} $ set to fit the CMB data
and $T_{\text{CMB}}=2.726$ K. 
We see that effects of the bounce lead to suppression of the low multipoles in CMB, what is 
favoured observationally. This possibility was indicated earlier in Ref. \cite{Piao:2003zm,Mielczarek:2008pf}.

Value of parameter $\mathcal{A}^2_{\mathcal{R}}$ can be calculated from the slow-roll inflation model
\begin{equation}
\mathcal{A}^2_{\mathcal{R}} = \frac{1}{2m^2_{\text{Pl}} \epsilon} \left( \frac{H}{2\pi} \right)^2 
\end{equation}
where $\epsilon$ is slow-roll parameter
\begin{equation}
\epsilon := \frac{1}{2\kappa} \left( \frac{1}{V}\frac{dV}{d\bar{\varphi}}\right)^2 
= \frac{1}{4\pi} \frac{m^2_{\text{Pl}}}{\bar{\varphi}^2}. 
\end{equation}
Based on this we can calculate value of the Hubble factor at the beginning of inflation
\begin{equation}
H_0 = \frac{\sqrt{2\pi \mathcal{A}^2_{\mathcal{R}}} \ m^2_{\text{Pl}}}{\bar{\varphi}_{\text{max}}} = 
3.8 \cdot 10^{-5} m_{\text{Pl}} \label{H0}
\end{equation}
where in the last equality we assumed $\bar{\varphi}_{\text{max}}=3.4  m_{\text{Pl}}$.

Important limitation of the method based on the low multipoles in CMB comes 
from the so called \emph{cosmic variance}. It is purely statistical effect
which is significant at the horizontal scales. In case of CMB this corresponds
to the low multipoles regime. Relative uncertainly coming from the cosmic 
variance is given by  
\begin{equation}
\frac{\Delta C_l}{C_l} = \sqrt{\frac{2}{2l+1}}. 
\end{equation}
Taking for example $l=2$ we obtain relative uncertainty equal $0.63$. Therefore 
outcome of measurement is comparable with its uncertainty. This effect cannot be 
removed and impose substantial limitation on our approach to constraint  
quantum cosmological models.  

\subsection{Discussion}

In the previous subsection we showed that bounce can leads to observed suppression 
of the low CMB multipoles. Now we would like to discuss whether this scenario is 
realistic and does non require fine-tuning. At the beginning we would like to however 
mention one important adjustment. Namely observed scale of cut-off in CMB spectrum 
overlaps with present size of cosmic horizon. This property was indicated in Ref. 
\cite{Contaldi:2003zv}. Therefore there is an intriguing possibility that observed 
cut-off is due to unknown evolution of the super-horizontal modes. Therefore it is 
not related with the pre-inflationary dynamics. Such an explanation seems to be more 
likely. However model of this super-horizontal damping does not exist yet. 

It is indicated by the observations that present scale of cut-off $\lambda_*(t_0)$ is 
comparable with $D_{\star}$, $\lambda_*(t_0)\approx D_{\star}$. Therefore 
\begin{equation}
D_{\star} \approx \lambda_*(t_i)  \frac{a_0}{a_i}=\lambda_*(t_i) 
e^N\frac{T_{\text{GUT}}}{T_{\text{dec}}}(1+z_{\text{dec}})
\end{equation}
Because $\lambda_*(t_i)=1/H_0$, taking Eq. \ref{H0} and Eq. \ref{Ndef} we 
obtain 
\begin{equation}
2Ne^{2N} = \xi  \label{NLamb}
\end{equation}
where 
\begin{equation}
\xi = \frac{2D^2_{\star}(2\pi)^2m^2_{\text{Pl}} \mathcal{A}^2_{\mathcal{R}}}{(1+z_{\text{dec}})^2} 
\left( \frac{T_{\text{dec}}}{T_{\text{GUT}}} \right)^2.
\end{equation}
It is worth to note that Eq. \ref{NLamb} has a form of the Lambert equation $W(z)e^{W(z)}=z$ which
define Lambert W function, therefore $N=\frac{1}{2}W(\xi)$. In order to determinate parameter $\xi$
we take $z_{\text{dec}}\simeq 1070$, $T_{\text{dec}} \simeq 0.2 $ eV, $T_{\text{GUT}} \simeq 10^{14} $ GeV 
and $\mathcal{A}^2_{\mathcal{R}}=2.6 \cdot 10^{-9} $. Based on this we obtain $\xi=5.1 \cdot 10^{62}$ and 
subsequently $N = 69.7$. Now one can determinate second independent parameter of inflation e.g. $m$. 
We can easily derive equation 
\begin{equation}
m \simeq \sqrt{\frac{3}{2}\mathcal{A}^2_{\mathcal{R}}} \frac{2\pi}{N} m_{\text{Pl}}.
\end{equation}
which gives $m=5.6 \cdot 10^{-6} m_{\text{Pl}}$. As we see the obtained parameters of inflation
are fully consistent with these usually considered. The model is in the full agreement with the 
present observational facts. Moreover any fine-tuning is not needed to explain suppression of 
the low multipoles by the quantum cosmological effects.  

\section{Other observational probes of LQC} \label{QGProbes}

Beside the effect of suppression of the low multipoles also other 
potential probes of quantum cosmologies are available. In this 
section we review four possible approaches. 

\subsection{Polarisation of CMB}

In Sec. \ref{CMB} we have shown how the quantum cosmological effects can 
be related with the spectrum of CMB anisotropy. This method gives us one 
possible approach to constraint quantum cosmologies. However observations 
of CMB brings us much more information than only anisotropies of 
temperature. Another important measured quantity is polarisation of CMB 
radiation. This polarisation can be described by the spectrum, which 
depends on the primordial perturbations. While E-type polarisation depends
on the both scalar and tensor components of perturbations, the B-type 
polarisation depends only on the tensor component. Spectrum of the E-type
is already observed and can be used e.g. to constraint cut-off of the 
power spectrum. Recently such a study were performed in Ref. \cite{Mortonson:2009xk}.
Based on the joined anisotropy and polarisation data it was shown  that 
scale of cut-off in the power spectrum is $C=\frac{k_c}{10^{-4}\text{Mpc}^{-1}}<4.2$ 
at 95$\%$ confidence level. While the constraint based only on the polarisation 
gives $C<5.2$. This result show that polarisation is a good tool to constraint 
the cut-off in the power spectrum. In particular while the considerations 
based only on the CMB anisotropies indicate a cut-off, the joined 
anisotropy and polarisation data indicate limit on the cut-off. This is 
crucial while constraining quantum cosmologies based on cut-offs 
resulting them.   

The second, B-type polarisation, still remains unreachable observationally.
However there is presently a huge effort to detect it. In particular 
mission like PLANCK aims to detect this type of polarisation.
If the B-type polarisation will be measured then amplitude of the tensor power 
spectrum can be determined. In case of slow-roll inflation this amplitude is given by
\begin{equation}
\mathcal{A}^2_T =\frac{8}{m^2_{\text{Pl}}} \left( \frac{H}{2\pi} \right)^2. 
\end{equation}
Because $H^2\simeq \frac{\kappa}{3}\rho$, the measurements of the B-type polarisation
enable us to determine energy scale of inflation. Therefore absolute values of 
parameters $N$ and $m$ can be found. This give us also restriction on the 
pre-bounce initial condition, in particular on $P(t_0)$.

\subsection{Non-Gaussianity}

When different modes of perturbations interact one another then field 
becomes non-Gaussian.  This interaction can be produced by the potential
of the scalar field, higher order corrections in the perturbative 
expansion or by the quantum gravitational effects. 

In case of the Gaussian field, all its statistical properties are 
fully described by the two point function $\langle \varphi_{\bf k_1} \varphi^*_{\bf k_2}\rangle = 
\delta^{(3)}({\bf k_1-k_2}) \frac{2\pi^2}{k^3} \mathcal{P}_{\varphi}(k)$ where 
$\mathcal{P}_{\varphi}(k)$ is power spectrum given by Eq. \ref{DefPowerSpect}. 
In order to describe non-Gaussian effects it is necessary to consider higher 
order correlation functions. The first contribution of non-linearity comes in 
tree-point function
\begin{equation}
\langle \varphi_{\bf k_1} \varphi_{\bf k_2} \varphi_{\bf k_3} \rangle = 
\delta^{(3)}({\bf k_1+k_2+k_3})  P_3({\bf k_1,k_2,k_3})  
\end{equation}
where $P_3({\bf k_1,k_2,k_3})$ is called bispectrum. In case of the Gaussian 
field, this spectrum is equal zero. 

The primordial non-Gaussianity, if present, could affect also the spectrum of CMB 
anisotropies leading to its non-Gaussianity. Present observations indicate however
that CMB spectrum is nearly Gaussian. This give us constraint on the cosmological
models with a huge amount of nonlinearity. In particular, based on these observations,
some quantum cosmological models can be constrained or even excluded. For example 
preliminary studies on non-Gaussianity in LQC were performed in Ref. \cite{Mielczarek:2008qw}.
In these studies, non-Gaussianity is produced by the specific scalar field potential not 
by the quantum gravity effects itself. However, this model gives an example of 
non-Gaussianity production in the bouncing universe. It was shown that in this model,
non-Gaussianity is produced in the vicinity of the points $t_+$ and $t_-$,
where Hubble factor reach its maximal value. Another example of non-Gaussianity production
in the bouncing cosmology can be found in Ref. \cite{Cai:2009fn}. In this paper 
production of the non-Gaussianity at the matter bounce is considered and indicate 
that this form of non-Gaussianity can be potentially distinguished from this produced 
during the inflationary phase.

\subsection{Transplanckian modes} 

When the length of the mode of perturbation approaching the Planck scale then 
semi-classical approximation fails. The notion of continuous wave is missed and 
and fully quantum gravitational considerations have to be applied. Therefore 
these so called transplanckian modes cannot be studied with use of Quantum 
Field Theory on Curved Spaces as it was done in the present paper. 
More adequate would be application of Quantum Field Theory on Quantum Spaces as this studied 
in Ref. \cite{Ashtekar:2009mb}. 

\begin{figure}[ht!]
\centering
\includegraphics[width=8cm,angle=0]{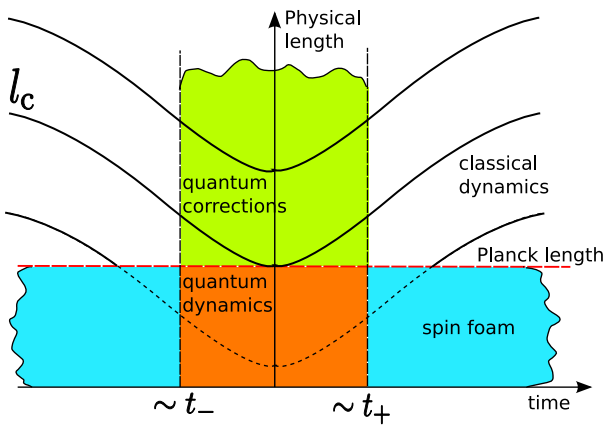}
\caption{Evolution of the different length scales in the bouncing universe.}
\label{ModesBounce}
\end{figure}

In Fig. \ref{ModesBounce} we show evolution of the different length scales in the bouncing universe. 
We see that modes with $\lambda>l_{\text{c}}$ can be described by the semi-classical approximation.
It is exactly the case considered in this paper. However when  $\lambda<l_{\text{c}}$ then new formulation
has to be applied. This would lead to potentially new effects. However these transplanckian modes
can decay before crossing the horizon during the inflation. Then any signature of the  
quantum gravity effects can be lost. However investigations as this performed in Ref. \cite{Danielsson:2002kx}
suggest that effects of transplanckian modes can lead to potentially observational effects.

\subsection{Large scale structures}

If the quantum cosmologies can give imprints on CMB, then some of these effects could be
seen also in the subsequent structures. The region of the low multipoles corresponds 
now to the largest visible distances in the Universe. Gravitational structures on these
scales are called large scale structure (LSS). Therefore observations of the large scale
structures are complementary to the observations of the low multipoles in CMB. These both 
methods were already applied to investigate effects of the bouncing cosmological scenario 
in Ref. \cite{Cai:2008qb}. In this paper authors predict oscillations in the power spectrum 
on the horizontal scales due to the bounce. However available observational data from e.g. SDSS 
are still not sufficient to probe these effects.

\section{Summary} \label{Summary}

In this paper we have shown possible way to relate quantum cosmological models 
with astronomical observations. This method can be used to constraint models of 
the universe in the Planck Epoch. In particular we constrained Barmero-Immirzi 
parameter $\gamma$, critical density $\rho_{\text{c}}$ and parameter $\lambda$.
However, present constraints are still very week 
and ambiguous. The part of this ambiguity can be removed by fixing the parameters 
of inflation ($N$ and $m$). Parameter N is today fixed by the requirement of solving 
the horizon problem. Then value of $m$ is determined from the amplitude of CMB 
anisotropies. However problem of horizon do not appear in the bouncing cosmologies. 
Therefore at present we have fixed only relation between $N$ an $m$ but not the 
absolute values. In order to fix one of these parameters, other observable has to 
be measured. The most promising is amplitude of tensor perturbations. These 
perturbations produce B-type polarisation in CMB. Therefore, if  this polarisation
would be measured then parameters $N$ and $m$ can be fixed. This will enable us 
to perform less ambiguous constraint on the LQC parameters, in particular on 
$\rho_\text{c}$. 

We have shown that phase of a bounce set proper initial conditions for inflation. 
Moreover parameters of inflation depends only logarithmically on the pre-bounce 
initial conditions. Subsequently we considered model of perturbations at the 
bounce and at the joined bounce+inflation phase. We showed that this second model can
give explanation of the suppression of the low multipoles in CMB. Moreover 
we have indicated that model is fully consistent and scale of the cut-off agree with the 
present size of horizon. This indicate that fine-tuning is not required to 
produce suppression on the horizontal scales. 
  
Beside the cosmological approach to constraint quantum gravity effects also 
other indirect methods are available. In particular astrophysical measurements 
of the Lorentz symmetry violation. It is in principle possible to derive quantitative 
predictions about this effect directly from LQG. However process of derivation
require construction of the semi-classical states, where unknown phenomenological 
parameters appear. Since their values are unknown, predictive power of this approach 
is marginal. If these difficulties would be removed then this method can be 
complementary with the cosmological approach. Also the quantum effects on the 
gravitational collapses gives possible way to constraint the LQG. Here the results 
are less ambiguous, however it is harder to relate them with some astrophysical data.

As we see the difficulties lies here on both theoretical and empirical side.
Without knowing the relation between the LQC and LQG we can treat parameters 
of this first rather as phenomenological. If these difficulties would not be 
overcome it will be hard to perform complementary constraints of the same parameters, 
based on different methods, e.g. observations of CMB and gamma ray bursts. 
 
\begin{acknowledgments}
Author is very grateful to A.~Barrau, T.~Cailleteau, J.~Grain and A.~Gorecki for 
discussion and hospitality during the visit in Grenoble where this paper was completed.
\end{acknowledgments}

\end{document}